\documentclass[preprint, superscriptaddress, preprintnumbers,amsmath,amssymb]{revtex4}

\allowdisplaybreaks
\allowdisplaybreaks[4]

\usepackage{graphicx}% Include figure files
\usepackage{dcolumn}% Align table columns on decimal point
\usepackage{bm}% bold math

\usepackage[dvipdfm,
            pdfstartview=FitH,
            bookmarks=true,
            bookmarksnumbered=true,
            bookmarksopen=true,
            colorlinks=true,
            pdfborder=001,
            linkcolor=blue,
            anchorcolor=blue,
            citecolor=blue,
            ]{hyperref}

\begin{document}

\title{Effective field theory for triaxially deformed nuclei}

\author{Q. B. Chen}\email{qbchen@pku.edu.cn}
\affiliation{Physik-Department, Technische Universit\"{a}t
M\"{u}chen, D-85747 Garching, Germany}

\affiliation{State Key Laboratory of Nuclear Physics and Technology,
School of Physics, Peking University, Beijing 100871, China}

\author{N. Kaiser}\email{nkaiser@ph.tum.de}
\affiliation{Physik-Department, Technische Universit\"{a}t
M\"{u}chen, D-85747 Garching, Germany}

\author{Ulf-G. Mei{\ss}ner}\email{meissner@hiskp.uni-bonn.de}
\affiliation{Helmholtz-Institut f\"{u}r Strahlen- und Kernphysik
and Bethe Center for Theoretical Physics,
Universit\"{a}t Bonn, D-53115 Bonn, Germany}

\affiliation{Institute for Advanced Simulation, Institut f\"{u}r Kernphysik,
J\"{u}lich Center for Hadron Physics and JARA-HPC, Forschungszentrum J\"{u}lich,
D-52425 J\"{u}lich, Germany}

\author{J. Meng}\email{mengj@pku.edu.cn}
\affiliation{State Key Laboratory of Nuclear Physics and Technology,
             School of Physics, Peking University, Beijing 100871, China}%
\affiliation{School of Physics and Nuclear Energy Engineering,
             Beihang University, Beijing 100191, China}%
\affiliation{Department of Physics, University of Stellenbosch,
             Stellenbosch, South Africa}%

\date{\today}

\begin{abstract}

Effective field theory (EFT) is generalized to investigate the rotational
motion of triaxially deformed even-even nuclei. A Hamiltonian, called the
triaxial rotor model (TRM), is obtained up to next-to-leading order (NLO)
within the EFT formalism. Its applicability is examined by comparing with a
five-dimensional collective Hamiltonian (5DCH) for the description of
the energy spectra of the ground state and $\gamma$ band in Ru isotopes.
It is found that by taking into account the NLO corrections, the ground state
band in the whole spin region and the $\gamma$ band in the low spin region
are well described. The results presented here indicate that it should be possible
to further generalize the EFT to triaxial nuclei with odd mass number.

\end{abstract}

\maketitle

%%%%%%%%%%%%%%%%%%%%%%%%%%%%%%%%%%%%%%%%%%%%%%%%%%%%%%%%%%
%                    begin  introduction
%%%%%%%%%%%%%%%%%%%%%%%%%%%%%%%%%%%%%%%%%%%%%%%%%%%%%%%%%%

\section{Introduction}\label{sec1}

As a quantum-mechanical complex many-body system, the atomic nucleus
exhibits modes of collective motion which have attracted much attention already  since the fifties
of the last century. The lowest-lying collective excitation
is the rotational mode. It is  well known that the existence
of nuclear rotation is due to the spontaneous  breaking~\footnote{Of course, there is no
spontaneous symmetry breaking in a finite system. This notion should  be understood
as an emergent symmetry that undergoes a spontaneous breaking as the system becomes
infinitely large. We still use the jargon often employed in nuclear physics.}
of the rotational
symmetry in the intrinsic frame of the nucleus, leading to the
appearance of deformation and thus  the nucleus itself has distinct
orientations. In the nuclear chart, the rare-earth and the actinide
nuclei comprise the typical mass regions where a large number of rotational
bands are observed.

In theoretical approaches, collective nuclear motions are mainly described by
collective geometric models, such as the Bohr-Mottelson model~\cite{Bohr1975},
or  the interacting boson model
(IBM)~\cite{Iachello1987book}. These models have led to significant achievements
due to their deep physical insights and mathematical beauty in the
description of collective nuclear excitations. They grasp the main features
(leading order effects, LO) of the rotational mode very well,  such as the fact that
rotational bands are built on certain vibrational excitation states.
However, as pointed out in Ref.~\cite{Papenbrock2011NPA},
it is very difficult to systematically extend such approaches, and hence they often fail
to account quantitatively for finer details (next-to-leading effects, NLO),
such as the change of the moment of inertia with spin. In addition,
it is also difficult  to compute results with reliable error estimates and thus to quantify the
limitations of these models.

To overcome the above mentioned deficiencies of  the traditional
approaches, Papenbrock and collaborators have presented a series of
works  in which effective field theory (EFT) is applied to describe
rotational and vibrational excitations of deformed nuclei.  Since
the initial paper in 2011~\cite{Papenbrock2011NPA} they have
completed a series of further works in Refs.~\cite{J.L.Zhang2013PRC,
Papenbrock2014PRC, Papenbrock2015JPG, Papenbrock2016PS,
Perez2015PRC, Perez2015PRC_v1, Perez2016PRC}. EFT is a theory based
on symmetry principles alone, and it exploits the separation of
scales for the systematic construction of the Hamiltonian
supplemented by a power counting. In this way, an increase in the
number of parameters (i.e., low-energy constants that need to be
adjusted to data) goes hand in hand with an increase in precision
and thereby counter balances the partial loss of predictive power.
Actually, EFT often exhibits an impressive efficiency as highlighted
by analytical results and economical means of calculations. In
recent decades, chiral effective field theory has enjoyed
considerable successes in low-energy hadronic and nuclear structure.
Pertinent examples include the descriptions of the nucleon-nucleon
interaction~\cite{Kolck1994PRC, Epelbaum2009RMP, X.L.Ren2016arXiv},
halo nuclei~\cite{Bertulani2002NPA, Hammer2011NPA, Ryberg2014PRC},
and few-body systems~\cite{Bedaque2002ARNPS, Griesshammer2012PPNP,
Hammer2013RMP}.

Through the application of EFT to deformed nuclei, the finer details mentioned
above can be properly addressed~\cite{Papenbrock2011NPA, J.L.Zhang2013PRC, Papenbrock2014PRC,
Papenbrock2015JPG, Papenbrock2016PS}. The uncertainties of the theoretical model
can be quantified~\cite{Perez2015PRC_v1}, and a consistent treatment of
currents together with the Hamiltonian is obtained~\cite{Perez2015PRC}. Let us
note that all of these investigations were concerned with axially deformed
nuclei.

The triaxial deformation of nuclei has been a subject of much
interest in the theoretical study of nuclear structure. Triaxial
deformation  is related to many interesting phenomena including the
$\gamma$ band~\cite{Bohr1975}, signature
inversion~\cite{Bengtsson1984NPA}, anomalous signature
splitting~\cite{Hamamoto1988PLB}, the wobbling
motion~\cite{Bohr1975}, chiral rotational
modes~\cite{Frauendorf1997NPA} and multiple chiral doublet
(M$\chi$D) bands~\cite{J.Meng2006PRC}. In particular, the wobbling
motion and chiral rotational modes are regarded as unique
fingerprints of stable triaxial nuclei.

Considering the successes and merits of EFT in the description of nuclear rotations
and vibrations, it would be interesting to extend it to the description of
triaxial nuclei. In this paper, as a first step, the EFT is adopted to construct
the Hamiltonian for the triaxial rigid rotor. The pertinent Hamiltonian is obtained up to
next-to-leading order (NLO). Taking the energy spectra of ground states and the
$\gamma$ bands in Ru isotopes as examples, the applicability of the EFT to
triaxial nuclei is examined.

The paper is organized as follows. In Sec.~\ref{sec2}, the EFT for triaxial
nuclei is constructed. The numerical details are introduced in Sec.~\ref{sec3} and in Sec.~\ref{sec4} the
results of the calculations are presented and discussed in detail. Finally, a summary
is given in Sec.~\ref{sec5} together with a perspective for future research directions.

%%%%%%%%%%%%%%%%%%%%%%%%%%%%%%%%%%%%%%%%%%%%%%%%%%%%%%%%%%
%                    begin  framework
%%%%%%%%%%%%%%%%%%%%%%%%%%%%%%%%%%%%%%%%%%%%%%%%%%%%%%%%%%

\section{Theoretical framework}\label{sec2}

In the effective field theory (EFT), the symmetry is (typically) realized nonlinearly, and the
Nambu-Goldstone fields parametrize the coset space
$\mathcal{G}/\mathcal{H}$, where $\mathcal{G}$ is the
symmetry group of the Hamiltonian, and $\mathcal{H}$, the symmetry group of the ground state,
is a proper subgroup of $\mathcal{G}$. The effective Lagrangian is built
from those invariants that are constructed from the fields in the coset space.
In the previous version of the EFT for nuclear rotation~\cite{Papenbrock2011NPA}, the
authors focussed on deformed nuclei with axial symmetry. In that case the nuclear
ground state is invariant under SO(2) rotations about the body-fixed symmetry axis,
while SO(3) symmetry is broken by the deformation. As a
consequence the Nambu-Goldstone modes belong to the two-dimensional coset
space $S^2=\textrm{SO}(3)/\textrm{SO}(2)$.
In this work, we consider triaxial nuclei for which SO(2) symmetry is further
broken by the loss of the axial symmetry and one is left with the (abelian)
discrete symmetry $\textrm{D}_2=\textrm{Z}_2\times \textrm{Z}_2$ (with four elements).
Hence, the Nambu-Goldstone modes lie on the three-dimensional coset space
SO(3)/$\textrm{D}_2$~\cite{Coleman1969PR, Callan1969PR,
Brauner2010Symmetry}.

As in Ref.~\cite{Papenbrock2014PRC}, we introduce the Nambu-Goldstone modes
as classical fields that are quantized later. We write these fields
in the space-fixed coordinate frame, where the three generators of infinitesimal
rotations about the space-fixed $x$, $y$, and $z$-axes are $J_x$, $J_y$, and
$J_z$, respectively. A triaxial nucleus is
invariant under $\textrm{D}_2$ rotations about the body-fixed
$x^\prime$, $y^\prime$, and $z^\prime$ axes with an angle $\pi$, while
SO(3) symmetry is broken by the deformation. The modes depend on the time-dependent Euler angles $\alpha(t)$,
$\beta(t)$, and $\gamma(t)$ which parametrize the unitary
transformations $U(\alpha,\beta,\gamma)$ related to SO(3) rotations in the following way:
\begin{align}
 U(\alpha,\beta,\gamma)
  =\exp\{-i\alpha(t)J_z\}\exp\{-i\beta(t)J_y\}
  \exp\{-i\gamma(t)J_z\}.
\end{align}
Note that the purely time-dependent variables $\alpha(t)$, $\beta(t)$, and
$\gamma(t)$ correspond to the zero modes of the system. They  parametrize rotations of the
deformed nucleus and  upon quantization they generate the rotational bands. Apparently,
one is dealing here with a  field theory in zero space-dimensions, i.e., ordinary quantum mechanics.

The underlying power counting is specified by
\begin{align}
 \alpha,\beta,\gamma \sim \mathcal{O}(1), \quad
 \dot{\alpha}, \dot{\beta}, \dot{\gamma} \sim \xi,
\end{align}
where the small parameter  $\xi$ denotes the energy scale of the rotational motion and the dot refers
to a time derivative.

\subsection{Effective Lagrangian}

The effective Lagrangian is built from invariants. These are constructed
from the components $a_t^x$, $a_t^y$, and $a_t^z$ of the angular velocity
arising from the decomposition
\begin{align}
 U^{-1}i\partial_t U=a_t^xJ_x+a_t^yJ_y+a_t^zJ_z.
\end{align}
By taking appropriate traces of the matrix-exponentials, the expansion
coefficients read
\begin{align}
 a_t^x &=-\dot{\alpha}\sin\beta\cos\gamma+\dot{\beta}\sin\gamma,\\
 a_t^y &=\dot{\alpha}\sin\beta\sin\gamma+\dot{\beta}\cos\gamma,\\
 a_t^z &=\dot{\alpha}\cos\beta+\dot{\gamma}.
\end{align}
One recognizes that these are the components of the angular velocity of the nucleus
in the body-fixed frame, according to rigid-body kinematics.

Considering  time-reversal invariance, the time derivatives in  $a_t^x$, $a_t^y$,
and $a_t^z$ allow only for even powers of the angular velocities.  The
quadratic invariants lead to the leading order (LO) Lagrangian
\begin{align}
 \mathcal{L}_{\textrm{LO}}&=\frac{1}{2}\mathcal{J}_1(a_t^x)^2
  +\frac{1}{2}\mathcal{J}_2(a_t^y)^2+\frac{1}{2}\mathcal{J}_3(a_t^z)^2,
\end{align}
where $\mathcal{J}_k$ ($k=1$, $2$, $3$) are parameters of order one to
be determined from experimental data.  It will become clear soon that these parameters are
equal to the moments of inertia about the three principal axes. Clearly, the LO effective
Lagrangian is of order $\xi^2$.

From the Lagrangian, one obtains the canonical momenta as:
\begin{align}
 p_\alpha &=\frac{\partial \mathcal{L}_{\textrm{LO}}}{\partial \dot{\alpha}}=-\mathcal{J}_1 a_t^x \sin\beta\cos\gamma
          +\mathcal{J}_2 a_t^y  \sin\beta\sin\gamma
         +\mathcal{J}_3 a_t^z  \cos\beta,\\
 p_\beta &=\frac{\partial \mathcal{L}_{\textrm{LO}}}{\partial \dot{\beta}}
          =\mathcal{J}_1 a_t^x  \sin\gamma
          +\mathcal{J}_2 a_t^y  \cos\gamma,\\
 p_\gamma&=\frac{\partial \mathcal{L}_{\textrm{LO}}}{\partial \dot{\gamma}}
          =\mathcal{J}_3 a_t^z.
\end{align}

\subsection{Effective Hamiltonian}

Using a  Legendre transformation, the Hamiltonian is given by
\begin{align}\label{eq8a}
 \mathcal{H}_{\textrm{LO}}
  &=\dot{\alpha}p_\alpha+\dot{\beta}p_\beta+\dot{\gamma}p_\gamma
  -\mathcal{L}_{\textrm{LO}}\notag\\
  &=\frac{1}{2\mathcal{J}_1}\Big(-p_\alpha\frac{\cos\gamma}{\sin\beta}
  +p_\beta\sin\gamma+ p_\gamma \cos\gamma\cot\beta \Big)^2\notag\\
 &\quad +\frac{1}{2\mathcal{J}_2}\Big(p_\alpha \frac{\sin\gamma}{\sin\beta}
 +p_\beta \cos\gamma -p_\gamma \sin\gamma\cot\beta \Big)^2 +\frac{1}{2\mathcal{J}_3}(p_\gamma)^2.
\end{align}
Noting that the expressions in the brackets are the three components
of angular momentum $I_1$, $I_2$, and $I_3$~\cite{Ring1980book},
\begin{align}
\label{eq4}
 I_1&=-p_\alpha \frac{\cos\gamma}{\sin\beta}+p_\beta \sin\gamma+ p_\gamma
  \cos\gamma\cot\beta,\\
\label{eq5}
 I_2&=p_\alpha\frac{\sin\gamma}{\sin\beta}+p_\beta\cos\gamma -p_\gamma
  \sin\gamma\cot\beta ,\\
\label{eq6}
 I_3&=p_\gamma,
\end{align}
we finally arrive at the Hamiltonian of triaxial rotor~\cite{Bohr1975}
\begin{align}
 \mathcal{H}_{\textrm{LO}}=\frac{I_1^2}{2\mathcal{J}_1}
 +\frac{I_2^2}{2\mathcal{J}_2}+\frac{I_3^2}{2\mathcal{J}_3}.
\end{align}
According to this formula, the physical interpretation of
$\mathcal{J}_k$ ($k=1$, $2$, $3$) as the moments of inertia  about
the three principal axes is obvious.

\subsection{Next-to-leading order}

At next-to-leading order (NLO), higher derivatives of the Nambu-Goldstone modes
appear, and we have to include terms of order $\xi^4$. Considering the behavior
of $a_t^x$, $a_t^y$, and $a_t^z$ under the discrete $\text{D}_2$ rotations and
demanding time-reversal invariance, three additional
terms can enter the effective Lagrangian up to fourth order
\begin{align}
 \mathcal{L}_{\textrm{NLO}} &=\mathcal{L}_{\textrm{LO}}+\Delta \mathcal{L}_{\textrm{NLO}},\\
 \Delta \mathcal{L}_{\textrm{NLO}} &=\frac{\mathcal{M}_1}{4}(a_t^x)^4
  +\frac{\mathcal{M}_2}{4}(a_t^y)^4+\frac{\mathcal{M}_3}{4}(a_t^z)^4,
\end{align}
where $\mathcal{M}_k$ ($k=1$, 2, 3) are parameters of order one
to be determined from the experimental energy spectra. Here, the
mixed terms are not included as $\mathcal{M}_k$ as well as $\mathcal{J}_k$
are defined with respect to the three principal axes.

Now, the canonical momenta are calculated as
\begin{align}
\label{eq1}
 p_\alpha
  &=\frac{\partial \mathcal{L}_{\textrm{NLO}}}{\partial \dot{\alpha}}
   =-a_t^x \sin\beta\cos\gamma
   [\mathcal{J}_1+\mathcal{M}_1(a_t^x)^2]+a_t^y\sin\beta\sin\gamma
   [\mathcal{J}_2+\mathcal{M}_2(a_t^y)^2]\notag\\
   & \qquad\qquad\qquad +a_t^z\cos\beta
   [\mathcal{J}_3+\mathcal{M}_3(a_t^z)^2],\\
\label{eq2}
  p_\beta
  &=\frac{\partial \mathcal{L}_{\textrm{NLO}}}{\partial \dot{\beta}}
   =a_t^x\sin\gamma
   [\mathcal{J}_1+\mathcal{M}_1(a_t^x)^2]+a_t^y\cos\gamma
   [\mathcal{J}_2+\mathcal{M}_2(a_t^y)^2],\\
\label{eq3}
  p_\gamma
  &=\frac{\partial \mathcal{L}_{\textrm{NLO}}}{\partial \dot{\gamma}}
   =a_t^z[\mathcal{J}_3
   +\mathcal{M}_3(a_t^z)^2].
\end{align}
Using the Legendre transformation, we obtain the next-to-leading order  Hamiltonian as
\begin{align}
 \mathcal{H}_{\textrm{NLO}}
  &=\dot{\alpha}p_\alpha+\dot{\beta}p_\beta+\dot{\gamma}p_\gamma
   -\mathcal{L}_{\textrm{NLO}}\notag\\
  &=\frac{1}{2}\mathcal{J}_1(a_t^x)^2
   +\frac{1}{2}\mathcal{J}_2(a_t^y)^2
   +\frac{1}{2}\mathcal{J}_3(a_t^z)^2
 +\frac{3}{4}\mathcal{M}_1(a_t^x)^4
   +\frac{3}{4}\mathcal{M}_2(a_t^y)^4
   +\frac{3}{4}\mathcal{M}_3(a_t^z)^4.
\end{align}

Combining Eqs.~(\ref{eq1})-(\ref{eq3}) and (\ref{eq4})-(\ref{eq6}),
one finds for the components of the angular momentum
\begin{align}
 I_1&=a_t^x[\mathcal{J}_1+\mathcal{M}_1(a_t^x)^2],\\
 I_2&=a_t^y[\mathcal{J}_2+\mathcal{M}_2(a_t^y)^2],\\
 I_3&=a_t^z[\mathcal{J}_3+\mathcal{M}_3(a_t^z)^2],
\end{align}
indicating that the corrections from the NLO terms generate principal  moments of inertia that
depend (quadratically) on the rotational frequency. The parameters  $\mathcal{M}_k$
($k=1, 2, 3$) are a measure of these non-rigidity effects. From the expression
for the angular momentum component $I_1$, one obtains its second and fourth power as:
\begin{align}
 I_1^2&=\mathcal{J}_1^2(a_t^x)^2\Big[1+2\Big(\frac{\mathcal{M}_1}{\mathcal{J}_1}\Big)(a_t^x)^2
       +\Big(\frac{\mathcal{M}_1}{\mathcal{J}_1}\Big)^2(a_t^x)^4\Big],\\
 I_1^4&=\mathcal{J}_1^4(a_t^x)^4\Big[1+4\Big(\frac{\mathcal{M}_1}{\mathcal{J}_1}\Big)(a_t^x)^2
       +6\Big(\frac{\mathcal{M}_1}{\mathcal{J}_1}\Big)^2(a_t^x)^4
      +4\Big(\frac{\mathcal{M}_1}{\mathcal{J}_1}\Big)^3(a_t^x)^6
       +\Big(\frac{\mathcal{M}_1}{\mathcal{J}_1}\Big)^4(a_t^x)^8\Big]~.
\end{align}
As the NLO terms are a correction to the LO ones, the ratio $\mathcal{M}_1/\mathcal{J}_1$
is expected to be small. Hence, in linear approximation we can use a combination of the last two equations
to express the terms proportional to (even powers of) $a_t^x$ in $\mathcal{H}_{\textrm{NLO}}$.
Setting
\begin{align}
 A_1I_1^2+B_1I_1^4=\frac{1}{2}\mathcal{J}_1(a_t^x)^2
  +\frac{3}{4}\mathcal{M}_1(a_t^x)^4,
\end{align}
one finds for the coefficients
\begin{align}
 A_1=\frac{1}{2\mathcal{J}_1}, \quad
 B_1=-\frac{\mathcal{M}_1}{4\mathcal{J}_1^4}~.
\end{align}
The analogous expressions in  $a_t^y$ and $a_t^z$ are written in terms of $I_2$ and $I_3$ with coefficients
\begin{align}
  A_2&=\frac{1}{2\mathcal{J}_2}, \quad
  B_2=-\frac{\mathcal{M}_2}{4\mathcal{J}_2^4},\\
  A_3&=\frac{1}{2\mathcal{J}_3}, \quad
  B_3=-\frac{\mathcal{M}_3}{4\mathcal{J}_3^4},
\end{align}
respectively. Therefore, the final Hamiltonian for the triaxial rotor
at NLO reads
\begin{align}
\label{eq8}
 \mathcal{H}_{\textrm{NLO}}
  &=\mathcal{H}_{\textrm{LO}}+\Delta \mathcal{H}_{\textrm{NLO}},\\
\label{eq7}
 \Delta \mathcal{H}_{\textrm{NLO}}
  &=-\frac{\mathcal{M}_1 I_1^4}{4\mathcal{J}_1^4}
   -\frac{\mathcal{M}_2 I_2^4}{4\mathcal{J}_2^4}
   -\frac{\mathcal{M}_3 I_3^4}{4\mathcal{J}_3^4}.
\end{align}

\subsection{Solutions of TRM Hamiltonian}

The TRM Hamiltonian at NLO $ \mathcal{H}_{\textrm{NLO}}$ in Eq.~(\ref{eq8}) is solved
by diagonalization. Since it is invariant under the discrete $D_2$ symmetry group, the basis
states $|\Psi_{IMK}\rangle$ can be chosen as~\cite{Bohr1975}
\begin{align}
|\Psi_{IMK}\rangle
=\frac{1}{\sqrt{2(1+\delta_{K0})}}\Big[|I,M,K\rangle &+(-1)^I|I,M,-K\rangle\Big], \quad K~\textrm{even},
\end{align}
where $|I,M,K\rangle$ denotes the Wigner $D$-functions (depending on the Euler angles $\alpha,\beta$ and $\gamma$). The
angular momentum projections onto the 3-axis in the intrinsic (body-fixed) frame
and the $z$-axis in the space-fixed frame
are denoted by $K$ and $M$, respectively. Note that $K\geq 0$ for even $I$,
while $K>0$ for odd $I$. Hence we have a total of $(I/2+1)$ basis states
for even $I$, and a total of $(I-1)/2$ basis states for odd $I$ with $I\geq 3$.

The calculations of the matrix elements of the Hamiltonian rely on the relations~\cite{Bohr1975}
\begin{align}
 I_+|I,M,K\rangle &=\sqrt{I(I+1)-K(K-1)}|I,M,K-1\rangle,\\
 I_-|I,M,K\rangle &=\sqrt{I(I+1)-K(K+1)}|I,M,K+1\rangle,\\
 I_3|I,M,K\rangle &=K|I,M,K\rangle,
\end{align}
where $I_+=I_1+iI_2$ and $I_-=I_1-iI_2$ are the raising and lowering operators of
angular momentum, respectively. The energy eigenvalues and eigenstates
for a given spin $I$ are obtained by solving the pertinent eigenvalue equation in matrix form.

\subsection{Impact of NLO corrections}

Before performing detailed calcualtions, let us study the impact of correction
term $\Delta \mathcal{H}_{\textrm{NLO}}$ of the NLO Hamiltonian in Eq.~(\ref{eq7})
to get an idea about the range of applicability of our EFT. For this purpose,
we assume moments of inertia of the irrotational type~\cite{Ring1980book}
\begin{align}\label{irro1}
\mathcal{J}_k &= \mathcal{J}_0 \sin^2(\gamma-2\pi k/3) , \\ \label{irro2}
\mathcal{M}_k &= \mathcal{M}_0 \sin^2(\gamma-2\pi k/3),\qquad (k=1,2,3),
\end{align}
setting the triaxial deformation parameter $\gamma = \pi/6 =30^\circ$ for simplicity.
Further, for estimating the energy spectra
as a function of spin $I$, the parameter $\mathcal{J}_0$ is taken as $25~\hbar^2/\textrm{MeV}$, while $\mathcal{M}_0$
is varied from $0$ to $25~\hbar^4/\textrm{MeV}^3$. Later, these parameters
will be fitted to experimental data.

\begin{figure}[!th]
  \begin{center}
    \includegraphics[width=9 cm]{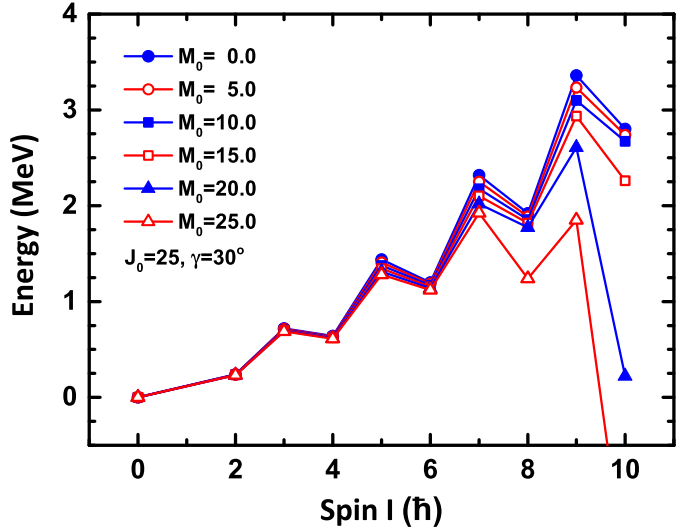}
    \caption{(Color online) Energy spectra as a function of spin $I$ calculated
    for the triaxial rotor model at NLO with parameters $\mathcal{J}_0=25~\hbar^2/\textrm{MeV}$,
$\mathcal{M}_0=(0\dots25)~\hbar^4/\textrm{MeV}^3$ and $\gamma=30^\circ$.\label{fig1}}
  \end{center}
\end{figure}

The obtained yrast energy spectra as a function of spin $I$ are shown in Fig.~\ref{fig1}.
One sees that with  increasing $\mathcal{M}_0$, the energy eigenvalues decrease. This feature is
due to the negative sign of $\Delta \mathcal{H}_{\textrm{NLO}}$. Because
$\Delta \mathcal{H}_{\textrm{NLO}}$ contains the fourth power of spin,
the impact of $\Delta \mathcal{H}_{\textrm{NLO}}$ for the high spin states
is larger than those for the low spin states. In addition, if
$\mathcal{M}_0$ is very large ($\mathcal{M}_0\geq 20~\hbar^4/\textrm{MeV}^3$),
the corrections from $\Delta \mathcal{H}_{\textrm{NLO}}$ lead to  irregular
energy spectra. In this case, the power counting is not longer obeyed, and
the limits of the EFT are exceeded.

\section{Numerical details}\label{sec3}

In the following, the newly developed EFT for the triaxial rotor is
applied to describe the experimental ground state and $\gamma$ bands
for the isotopes $^{102}$Ru up to $^{112}$Ru. The data are taken
from the compilation of the National Nuclear Data Center
(NNDC)~\footnote{http://www.nndc.bnl.gov/ensdf/.}. In the
calculations, both $\mathcal{J}_k$ and $\mathcal{M}_k$ are assumed
to be of the irrotational type, see
Eqs.~(\ref{irro1})~and~(\ref{irro2}). As a first strategy,
$\mathcal{J}_0$ and $\mathcal{M}_0$ are fitted to the experimental
energies of the lowest members of the ground state band, $0_1^+$,
$2_1^+$, $4_1^+$, $6_1^+$, while the triaxial deformation parameter
$\gamma$  is taken from the covariant density functional theory
(CDFT) calculations employing the effective interaction
PC-PK1~\cite{P.W.Zhao2010PRC}. In these constrained CDFT
calculations, the Dirac equation for a nucleon is solved in a
three-dimensional harmonic oscillator basis, which in the present
case includes 12 major oscillator shells. The pairing correlations
are treated within the BCS scheme utilizing a delta pairing-force.
The obtained energy spectra will be compared with the results of the
five-dimensional collective Hamiltonian (5DCH)~\cite{Niksic2011PPNP}
in order to examine the applicability of the present
EFT approach. %%\mathcal{J}_0\mathcal{J}_0

In the 5DCH calculations, both the collective potential and the inertial parameters
are calculated with the constrained CDFT using the effective interaction
PC-PK1~\cite{P.W.Zhao2010PRC}. In the calculations of the moments of
inertia $\mathcal{J}_k$, the Inglis-Belyaev formula is
used~\cite{Ring1980book}. It usually underestimates the experimental moments of inertia
due to the absence of the contributions from time-odd nuclear mean-fields and the absence of the
so-called Thouless-Valatin (TV) dynamical
rearrangement contributions~\cite{Z.P.Li2012PRC}. The proper inclusion of these effects
requires very demanding computations. In order to account for these in an
approximate way, one multiplies the moments of inertia (using the  Inglis-Belyaev formula)
with a fudge factor $f$, that is fitted to reproduce the energy of the  experimentalvalue of the  $2_1^+$ state.

As a second strategy, the three parameters $\mathcal{J}_0$, $\mathcal{M}_0$,
and $\gamma$ are fitted to the data. This corresponds to a genuine
EFT approach supplemented by the condition of irrotational moments of inertia,
which is common for the triaxially deformed nuclei considered in this work. We do not consider the
case where all the $\mathcal{J}_k$ and $\mathcal{M}_k$ are
determined from  a fit to the data. In that case, one would have to fit 3 (6) parameters at LO (NLO).

The results based on this approach are compared to those from the first
strategy to order to examine the quality of covariant density functional
theory (CDFT) in predicting triaxial deformations of nuclei.

%%%%%%%%%%%%%%%%%%%%%%%%%%%%%%%%%%%%%%%%%%%%%%%%%%%%%%%%%%
%                    begin  results and discussion
%%%%%%%%%%%%%%%%%%%%%%%%%%%%%%%%%%%%%%%%%%%%%%%%%%%%%%%%%%

\section{Results and discussion}\label{sec4}

\subsection{Quasi-particle alignment analysis}

\begin{figure}[!th]
  \begin{center}
    \includegraphics[width=11 cm]{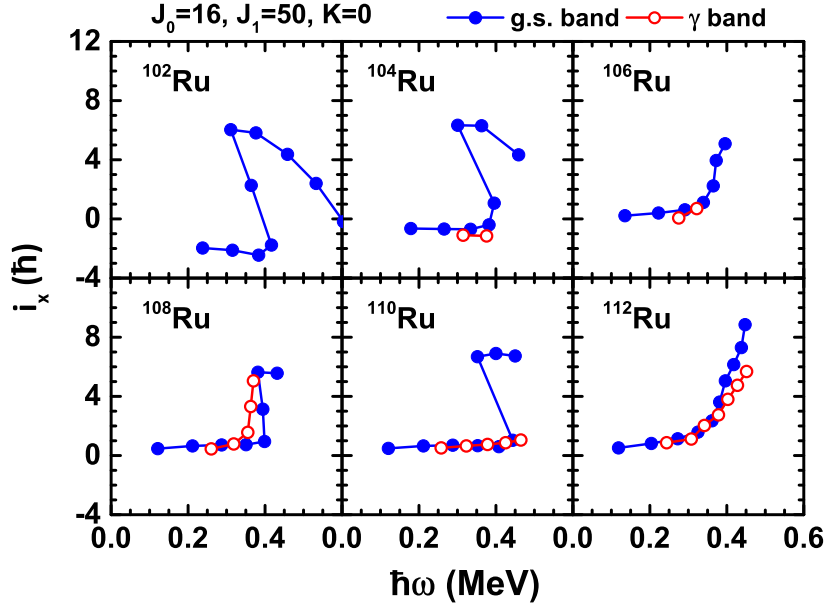}
    \caption{(Color online) Quasi-particle alignments of ground state and $\gamma$
    bands for the isotopes $^{102\textrm{-}112}$Ru.  The calculations at $K=0$
     were performed with Harris parameters  $\mathfrak{J}_0=16.0~\hbar^2/\textrm{MeV}$ and
    $\mathfrak{J}_1=50.0~\hbar^2/\textrm{MeV}^3$.}\label{fig2}
  \end{center}
\end{figure}

In order to establish the applicability of the triaxial rotor model (TRM), we study first the
quasi-particle alignment of the experimental energy spectra as described in
Ref.~\cite{Bengtsson1979NPA}. The quasi-particle alignment $i_x$  is defined as the difference
between the spin of the data and that of a reference rotor  (with moment of inertia
${\mathfrak J}_0+\omega^2 {\mathfrak J_1}$).
Such an empirical study provides information about the spin
at which a nucleon-pair breaks and thus the collective rotor model becomes inapplicable.
We consider the $K=0$ band and choose for the so-called
Harris parameters $\mathfrak{J}_0=16.0~\hbar^2/\textrm{MeV}$
and $\mathfrak{J}_1=50.0~\hbar^2\textrm{MeV}^3$, which  describe the dependence
of moments of inertia on the rotational frequency in the form $\mathfrak{J}_0+\omega^2 \mathfrak{J}_1$.
The obtained alignments $i_x$ as a function of the rotational frequency $\omega$ are shown in
Fig.~\ref{fig2}.
In all cases, the calculated alignments display a nearly constant behavior at
low $\hbar\omega$ (corresponding to spins $I\leq 10$) whereas  a drastic increase sets in
at  higher $\hbar\omega$. For this reason the range of applicability of  the TRM
and 5DCH Hamiltionians is restricted to the region with spins $I\leq 10$.
Therefore, we consider in the following only energy spectra in this low-spin region.

\subsection{$^{112}$Ru}

\begin{figure}[!t]
  \begin{center}
  \hspace{-2.5 cm}
    \includegraphics[width=9 cm]{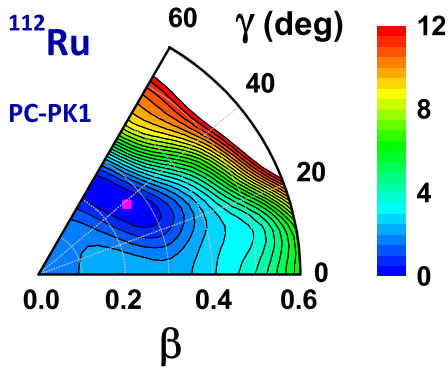}
    \vspace{-2.2 cm}
    \caption{(Color online) Contour lines of constant potential energy  in the
    $\beta\gamma$-plane for the ground-state configuration of $^{112}$Ru calculated in constrained
    covariant density functional theory with the effective interaction PC-PK1.
    Energies (in MeV) are measured with respect to the absolute minimum indicated by a square.
The energy separation between neighboring contour lines is 0.5~MeV. }\label{fig3}
  \end{center}
\end{figure}

In this subsection, the $^{112}$Ru nucleus is selected to demonstrate
the applicability of the EFT in the description of collective rotations of
triaxially deformed nuclei. In the TRM calculations, the triaxial deformation
parameter $\gamma$ as obtained from constrained CDFT calculations is used in
the first fit strategy. Fig.~\ref{fig3} shows
contour lines of constant potential energy in the $\beta\gamma$-plane as
obtained with the PC-PK1 effective interaction~\cite{P.W.Zhao2010PRC}. The
potential energy is measured with respect to its absolute minimum (marked
by a square in Fig.~\ref{fig3}). It is found that the ground
state of $^{102}$Ru has the
deformation parameters $\beta =0.26$ and $ \gamma =38.4^\circ$ as well as a
moderate $\gamma$-softness.

\begin{figure}[!th]
  \begin{center}
    \includegraphics[width=9 cm]{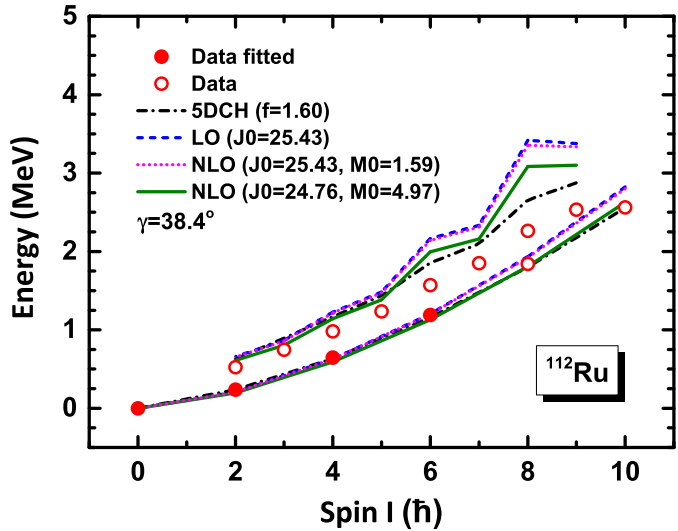}
    \caption{(Color online) Energy spectra for the ground state and $\gamma$
    band in $^{112}$Ru calculated in the triaxial rotor model at LO and NLO in comparison to
    results of the 5-dimensional collective Hamiltonian (5DCH).}\label{fig4}
  \end{center}
\end{figure}

With the given triaxial deformation parameter $\gamma =38.4^\circ$, the other
moment of inertia parameters $\mathcal{J}_0$ and $\mathcal{M}_0$ of the irrotational TRM are determined by fitting to
the experimental data. In the LO calculation, the value of $\mathcal{J}_0$ is $\mathcal{J}_0=
25.43~\hbar^2/\textrm{MeV}$. Fixing this value and performing the the
NLO calculation gives $\mathcal{M}_0=1.59~\hbar^4/\textrm{MeV}^3$ for the non-rigidity parameter.
If $\mathcal{J}_0$ and $\mathcal{M}_0$ are simultaneously fitted to the
data in an NLO calculation, one obtains somewhat different values,
$\mathcal{J}_0=24.76~\hbar^2/\textrm{MeV}$ and $\mathcal{M}_0=4.97~\hbar^4/\textrm{MeV}^3$.
The corresponding energy spectra of the ground state and $\gamma$ bands are shown in
Fig.~\ref{fig4} as a function of spin $I$ in comparison to the
results of the 5DCH Hamiltonian. One observes that that both the LO and NLO calculation
provide a reasonable description of the experimental data. In the LO calculation
the energies of ground states $8_1^+$ and $10_1^+$ are
somewhat overestimated and the description of the $\gamma$ band in high-spin region
$I\geq 6$ is not as good as for low spins. In the case of
NLO calculation with $\mathcal{J}_0$ fixed, the obtained results are very similar
to those at LO, since $\mathcal{M}_0$ is small and thus only small
corrections are involved.
In the case where $\mathcal{J}_0$ and $\mathcal{M}_0$ are simultaneously
fitted, the description of the data is somewhat improved. Nevertheless, there
are still some deviations for the high-spin states in the $\gamma$ band.
By comparing with the 5DCH calculation, which reproduces better the ground state and
$\gamma$ bands, one can attribute the deviations in the TRM to the neglect of
the vibrational degrees of freedom. Consequently, in the future one should
include systematically the vibrational degrees of freedom in the EFT.

\begin{figure}[!th]
  \begin{center}
    \includegraphics[width=9 cm]{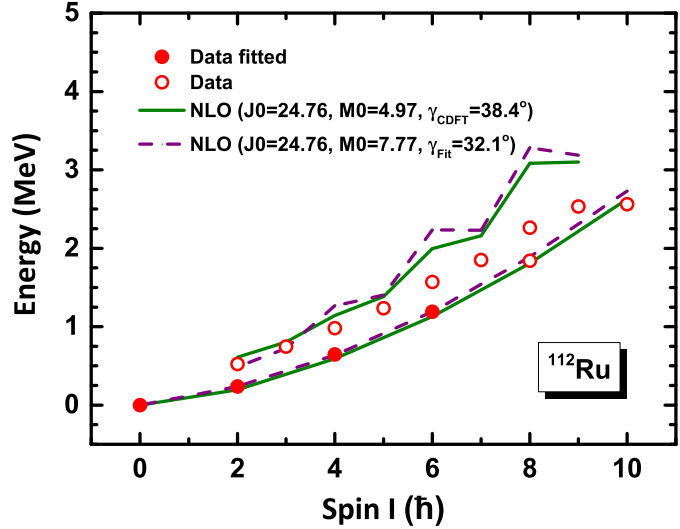}
    \caption{(Color online) Same as Fig.~\ref{fig4}. The results of both strategies
    are compared with each other (see the text for  details).}\label{fig8}
  \end{center}
\end{figure}

In Fig.~\ref{fig8}, we show the results obtained by the second
strategy, where $\mathcal{J}_0$, $\mathcal{M}_0$, and
$\gamma$ are fitted simultaneously to the data, and compare them with those of
the first strategy. The corresponding parameter values are
$\mathcal{J}_0=24.76~\hbar^2/\textrm{MeV}$, $\mathcal{M}_0=7.77~\hbar^4/\textrm{MeV}^3$,
and $\gamma=32.1^\circ$. Note that  $\gamma$
is close to the value given by the CDFT calculation, suggesting that the triaxial
deformation predicted by CDFT is quite reliable. From Fig.~\ref{fig8}, one observes
that the description of the ground state band is similar in both
strategies. The same feature applies
to the calculated $\gamma$ bands. It is comforting to see that the effective
field theory without any external input works equally well as the (more microscopic) CDFT calculation.

\subsection{Isotopes $^{102}$Ru up to  $^{110}$Ru}

After the successful description of $^{112}$Ru, one can perform analogous
calculations for the lighter isotopes $^{102}$Ru up to $^{110}$Ru. Such a
systematic study over a long chain of isotopes provides further tests of the
applicability of the EFT.

\begin{figure}[!ht]
  \begin{center}
    \includegraphics[width=9 cm]{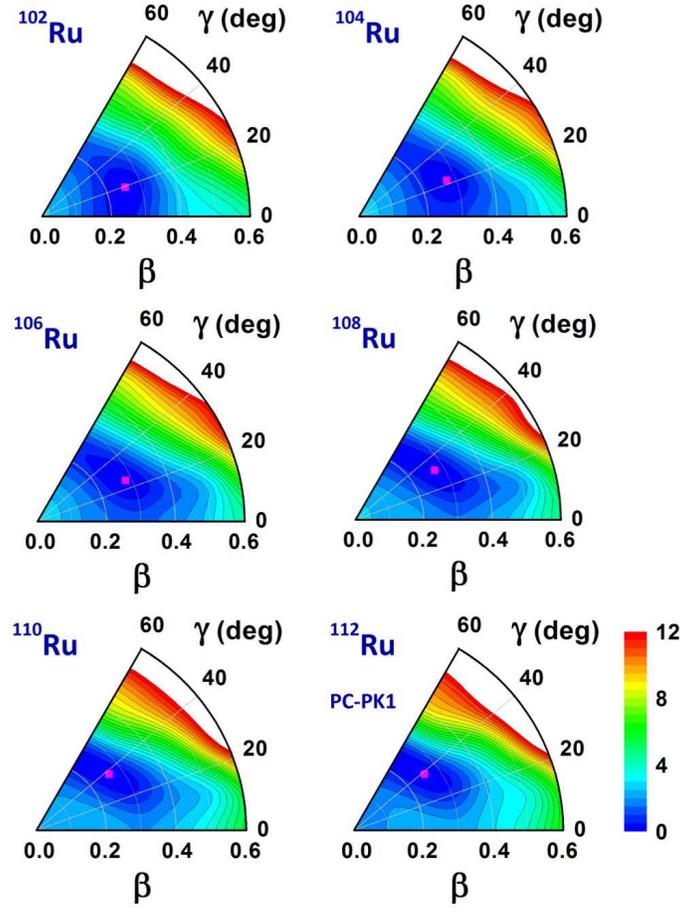}
    \caption{(Color online)  Contour lines of constant potential energy  in
    the $\beta\gamma$-plane for the ground-state configuration of the isotopes $^{102\textrm{-}112}$Ru.}\label{fig5}
  \end{center}
\end{figure}

In Fig.~\ref{fig5}, the contour lines of constant potential energy in the
$\beta\gamma$-plane are shown for $^{102}$Ru up to $^{112}$Ru. One observes that all potential energy surfaces
possess a triaxially deformed minimum and they exhibit softness along the
$\gamma$-direction. The sequence of plots in Fig.~\ref{fig5} shows that with increasing neutron
number the $\gamma$-coordinate of the minimum becomes larger. These angles $\gamma$ together
with the $\beta$-coordinate at the minimum are listed in Table~\ref{tab1}. One observes that
$\gamma$ doubles from $19.2^\circ$ to $38.4^\circ$ when the mass number ranges from 102
to 112. The values in Table~\ref{tab1} provide the input to the EFT calculations based on the first
fit strategy.

\begin{table}[!ht]
\caption{Deformation parameters $(\beta,\gamma)$ of the ground state in $^{102\textrm{-}112}$Ru
calculated with the constrained CDFT employing the PC-PK1 effective interaction.}
\label{tab1}
\begin{ruledtabular}
\begin{tabular}{cccc}
Nucleus    & $(\beta,\gamma)$   & Nucleus  & $(\beta,\gamma)$ \\
\colrule
$^{102}$Ru & $(0.25, 19.2^\circ)$ & $^{104}$Ru & $(0.28, 22.2^\circ)$ \\
$^{106}$Ru & $(0.28, 24.9^\circ)$ & $^{108}$Ru & $(0.27, 31.9^\circ)$ \\
$^{110}$Ru & $(0.26, 37.6^\circ)$ & $^{112}$Ru & $(0.26, 38.4^\circ)$ \\
\end{tabular}
\end{ruledtabular}
\end{table}

\begin{figure}[!ht]
  \begin{center}
    \includegraphics[width=12.5 cm]{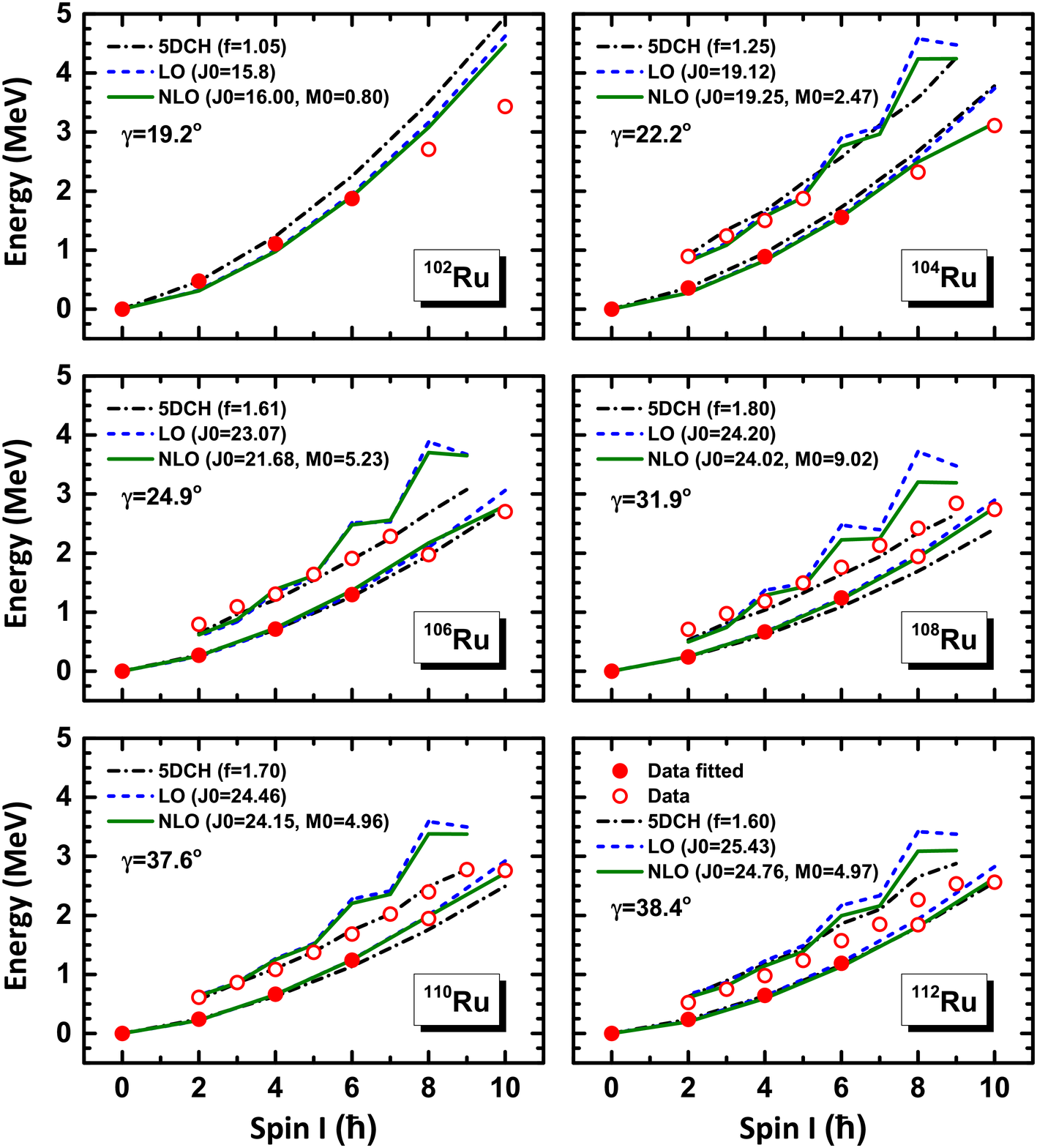}
    \caption{(Color online) Same as Fig.~\ref{fig4}, but for
$^{102}$Ru up to $^{112}$Ru.}\label{fig6}
  \end{center}
\end{figure}

In Fig.~\ref{fig6}, the energy spectra of the ground state and
$\gamma$ bands in the isotopes $^{102}$Ru up to $^{112}$Ru calculated
at LO and NLO are shown in comparison to experimental data and
results from the 5DCH Hamiltionian. We find similar results and draw
the same conclusions for these lighter Ru isotopes as for $^{112}$Ru.
Overall, the description at NLO is better than at LO, but there are
still some deviations between the NLO results and the data for the
high spin states in the $\gamma$ bands. Since the 5DCH results are
in good agreement with the data, this points again towards the
importance of including vibrational degrees of freedom in the EFT
formulation.

\begin{figure}[!t]
  \begin{center}
    \includegraphics[width=9 cm]{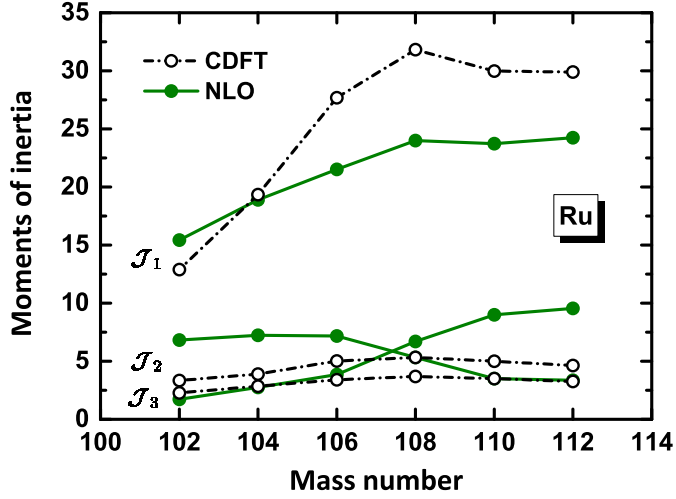}
    \caption{(Color online) Moments of inertia $\mathcal{J}_k$
($k=1$, 2, 3) entering the TRM in comparison to those computed with
constrained CDFT, using the triaxial deformation parameter $\gamma$
from CDFT (strategy one).}\label{fig7}
  \end{center}
\end{figure}

As we have mentioned, the inertial parameters of the 5DCH Hamiltonian
are calculated with the CDFT. In Fig.~\ref{fig7}, the three principal moments
of inertia $\mathcal{J}_{1,2,3}$ of the ground state are compared to
those determined at NLO over the mass region 102$-$112. One finds
appreciable differences and therefore the non-rigidity parameters
$\mathcal{M}_k$ ($k=1$, 2, 3) should also be extracted from
constrained CDFT calculations in the future. This way
all parameters of the triaxial rotor model (TRM) at NLO would be
determined in a fully microscopic manner.

\begin{figure}[!th]
  \begin{center}
    \includegraphics[width=12.5 cm]{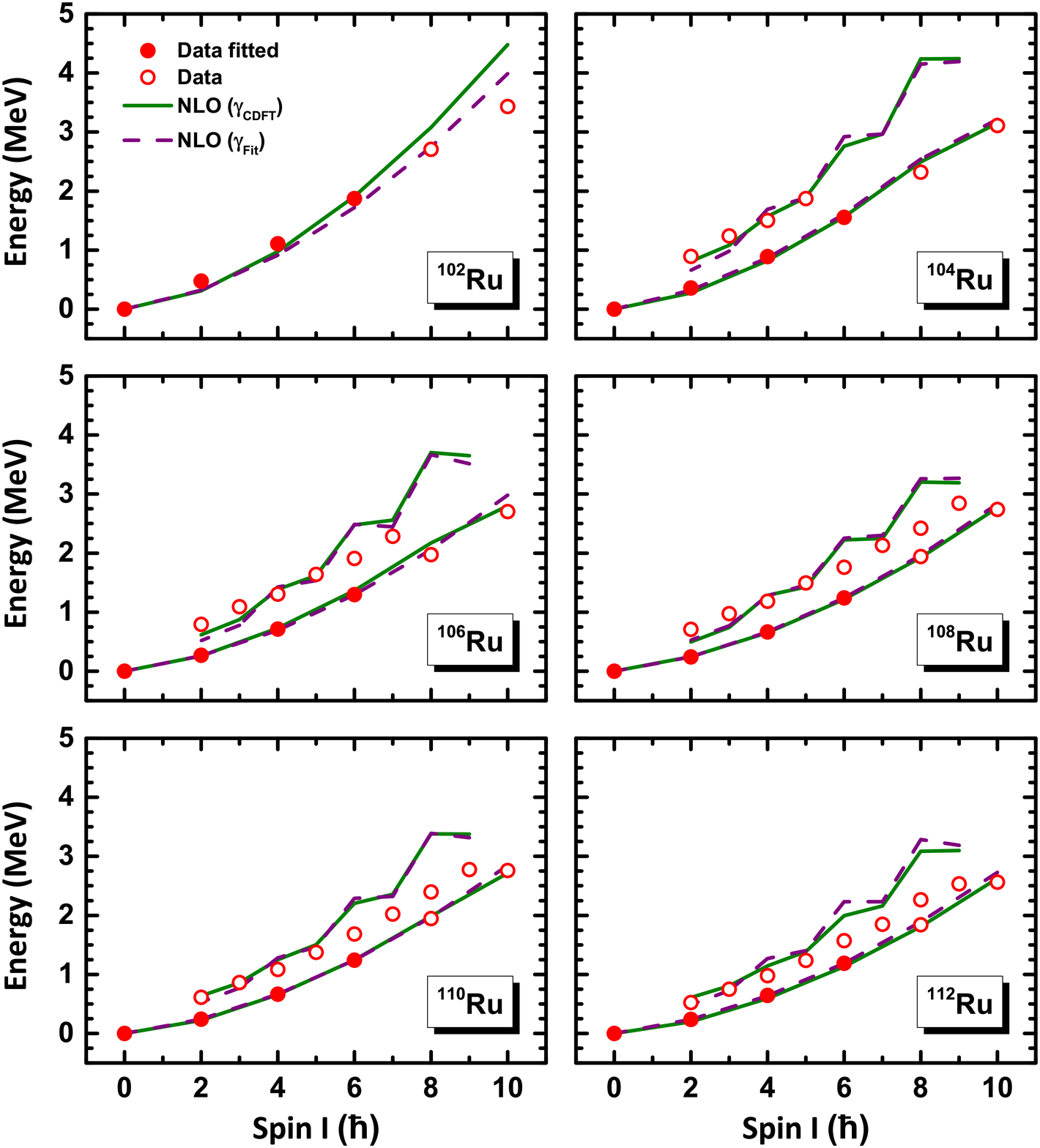}
    \caption{(Color online) Energy spectra for the ground state and $\gamma$-bands in $^{102}$Ru up to
 $^{112}$Ru at NLO taking the triaxial deformation parameter $\gamma$
    from the 5DCH calculation (solid green lines) or directly from the EFT fit
    (dashed purple lines).}\label{fig9}
  \end{center}
\end{figure}

Next, we consider the results of the genuine EFT approach, where all
pertinent parameters are determined from a fit to data. In
Tab.~\ref{tab2}, we give the resulting values for
$\mathcal{J}_0$, $\mathcal{M}_0$ and $\gamma$ for the considered Ru
isotopes considered in comparison to the values obtained in strategy
one. The two sets of parameters are similar, but there some
differences, most visible in the triaxial deformation parameter
$\gamma$.  It remains to be seen whether these differences persist
if the vibrational degrees of freedom are included in the EFT
formulation. The  energies in the ground state and the $\gamma$ band
at NLO are displayed in Fig.~\ref{fig9} for both fit strategies and
the resulting description of all data turns out to be very similar in
both approaches.

\begin{table}[!ht]
\caption{Parameters used in the NLO calculations for strategy one (I)
and two (II). The units of $\mathcal{J}_0$ and $\mathcal{M}_0$ are $\hbar^2/\textrm{MeV}$ and
$\hbar^4/\textrm{MeV}^3$, respectively.} \label{tab2}
\begin{ruledtabular}
\begin{tabular}{cccccccccccc}
 & Nucleus    & $^{102}$Ru & $^{104}$Ru & $^{106}$Ru & $^{108}$Ru & $^{110}$Ru & $^{112}$Ru \\
\colrule
   & $\gamma_{\textrm{CDFT}}$ & $19.2^\circ$ & $22.2^\circ$ & $24.9^\circ$  & $31.9^\circ$ & $37.6^\circ$ & $38.4^\circ$ \\
I  & $\mathcal{J}_0$          &  16.00       &  19.25       &  21.68        &  24.02       &  24.15       &  24.76 \\
   & $\mathcal{M}_0$          &  ~0.80       &  ~2.47       &  ~5.23        &  ~9.02       &  ~4.96       &  ~4.97 \\
\hline
   & $\gamma_{\textrm{Fit}}$  & $24.7^\circ$ & $28.0^\circ$ & $29.9^\circ$  & $33.6^\circ$ & $33.8^\circ$ & $32.1^\circ$ \\
II & $\mathcal{J}_0$          &  17.42       &  18.11       &  22.85        &  23.65       &  23.81       &  24.76 \\
   & $\mathcal{M}_0$          &  ~1.23       &  ~4.56       &  ~4.54        &  ~8.76       &  ~7.07       &  ~7.77 \\
\end{tabular}
\end{ruledtabular}
\end{table}

%%%%%%%%%%%%%%%%%%%%%%%%%%%%%%%%%%%%%%%%%%%%%%%%%%%%%%%%
%                    begin  summary
%%%%%%%%%%%%%%%%%%%%%%%%%%%%%%%%%%%%%%%%%%%%%%%%%%%%%%%%

\section{Summary}\label{sec5}

In this work the effective field theory for the collective
rotational motion has been  generalized to triaxially deformed
nuclei. The Hamiltonian of the triaxial rotor model has been
constructed up to next-to-leading order in the EFT power counting.
Taking the energy spectra of the ground state and $\gamma$ bands of
the even isotopes $^{102}$Ru up to $^{102}$Ru as benchmarks, the
applicability of the EFT has been examined by describing the
pertinent data for spins $I\leq 10$ and
by comparing to results obtained with a five-dimensional
collective Hamiltonian. It is found that the description at NLO
is overall better than at LO. Nevertheless, there are still some
deviations between the NLO calculation and the data for high-spin
states in the $\gamma$ bands. This points towards the importance of
including vibrational degrees of freedom in the EFT formulation.

In addition, we have compared two strategies of fitting parameters.
In the first strategy, $\gamma$ is taken from a CDFT calculation,
and in the second strategy, $\gamma$ is also fitted to the data (the
genuine EFT approach). The corresponding results show that
the EFT for collective nuclear rotation can be applied without
referring to any microscopic (model-dependent) input. We have found
that the value of $\gamma$ as obtained in the second strategy is
close to the one predicted by CDFT. This suggests that CDFT is a
reliable microscopic approach to calculate ground state properties.
Hence, we can (but need not) combine EFT and CDFT to describe the
rotational spectra of deformed nuclei.

The results presented give us a strong motivation to further
generalize the EFT for triaxially deformed nuclei with odd mass
number and to include systematically the vibrational degrees of
freedom.

\section*{Acknowledgements}

We thank P. Ring and W. Weise for helpful discussions. This work was
supported in part by the Deutsche Forschungsgemeinschaft (DFG) and
National Natural Science Foundation of China (NSFC) through funds
provided to the Sino-German CRC 110 ``Symmetries and the Emergence
of Structure in QCD'', the China Postdoctoral Science Foundation
under Grants No. 2015M580007 and No. 2016T90007, the Major State 973
Program of China (Grant No. 2013CB834400), and the NSFC under Grants
No. 11335002, No. 11375015, No. 11461141002, and No. 11621131001.
The work of UGM was also supported by the Chinese Academy of Sciences
(CAS) President's International Fellowship Initiative (PIFI)
(Grant No. 2017VMA0025).

%%%%%%%%%%%%%%%%%%%%%%%%%%%%%%%%%%%%%%%%%%%%%%%%%%%%%%%%
%                  begin refereee
%%%%%%%%%%%%%%%%%%%%%%%%%%%%%%%%%%%%%%%%%%%%%%%%%%%%%%%%

\end{document}